\documentclass[12pt]{article}

\usepackage{epsfig,amstext,amsthm, amsmath, amsfonts, a4, color,amssymb}
\usepackage{latexsym}
\usepackage{natbib}
\usepackage{graphicx}
\usepackage{placeins, bm}

\def\hang{\hangindent\parindent}
\def\rf{\par\noindent\hang}

\renewcommand{\mathbf}{\boldsymbol}

\renewcommand{\hat}[1]{\widehat{#1}}

\newcommand{\ba}{{\mathbf{a}}} 

\newcommand{\cc}{{\mathbf{c}}}

\newcommand{\II}{{\mathbf{I}}}

\newcommand{\XX}{{\mathbf{X}}}
\newcommand{\YY}{{\mathbf{Y}}}

\newcommand{\be}{{\mbox{\boldmath $\beta$}}}

\newcommand{\vare}{{\mbox{\boldmath $\varepsilon$}}}

\newcommand{\zer}{{\mbox{\boldmath $0$}}}

\begin{document}

\title{The Performance of the Turek-Fletcher Model Averaged Confidence Interval}
\author{Paul Kabaila$^1$, A.H. Welsh$^2$ \& Rheanna Mainzer$^1$}

\date{\today}

\maketitle

\abstract{ We consider the model averaged tail area (MATA) confidence interval proposed by Turek and Fletcher, {\sl CSDA}, 2012, in the simple situation in which we average over two nested linear regression models.
We prove that the MATA for {\sl any} reasonable weight function belongs to the class of confidence intervals defined by Kabaila and Giri, {\sl JSPI}, 2009. Each confidence interval in this class is specified by two functions $b$ and $s$. Kabaila and Giri show how to compute these functions so as to optimize these intervals in terms of satisfying the coverage constraint and
minimizing the expected length for the simpler model, while ensuring that the expected length has desirable properties for the full model. These Kabaila and Giri ``optimized'' intervals provide an upper bound on the performance of the MATA for an arbitrary weight function. This fact is used to evaluate the MATA for a broad class of weights based on exponentiating a criterion related to Mallows' $C_P$. Our results show that, while far from ideal, this MATA performs surprisingly well, provided that we choose a member of this class that does not put too much weight on the simpler model.
}

\noindent {\it Keywords:} Akaike Information Criterion (AIC); confidence interval; coverage probability; expected length; model selection; nominal coverage, regression models.


\noindent
\textit{$^1$Department of Mathematics and Statistics,
        La Trobe University, Victoria 3086, Australia\\
        $^2$Mathematical Sciences Institute,
        The Australian National University, Canberra ACT 0200, Australia }

\newpage

\section{Introduction}

\medskip

The Turek  and Fletcher (2012) model averaged tail area confidence interval (MATA) endpoints are obtained by solving a weighted average of the tail area equations for the confidence interval endpoints for each model.  Turek  and Fletcher first consider giving a weight to each model by exponentiating AIC/2, where AIC is the Akaike Information Criterion for the model. This is the earliest form of weight used in frequentist model averaging and was first proposed by Buckland \textit{et al.} (1997).  Turek  and Fletcher also consider MATA for weights obtained by replacing AIC by either the Akaike Information Criterion corrected for small samples, AIC$_c$, or the Bayesian Information Criterion, BIC.  They provide some comparison of the performance of MATA for these three weights in particular examples using simulation.

We evaluate the performance of the MATA, with nominal coverage $1-\alpha$ and specified weight function, by focusing on the coverage and the scaled expected length, where the expected length is scaled with respect to the length of the standard confidence interval (based on the full model) with coverage equal to the minimum coverage probability of the MATA.  We consider a simple situation in which we average over a linear regression model with independent and identically distributed normal errors (${\cal M}_2$) and the same model with a linear constraint on the regression parameters (${\cal M}_1$).  Kabaila, Welsh and Abeysekera (2015) derive computationally convenient, exact expressions for the coverage probability and the scaled expected length of the MATA, so that we can readily obtain highly-accurate numerical results without resorting to simulations.  In the same simple situation, we consider MATA  with {\sl any} reasonable weight function. We prove that the MATA with {\sl any} reasonable weight function belongs to a subclass of the class of confidence intervals, denoted by $J(b,s)$, defined by Kabaila and Giri (2009). Each confidence interval in this class is specified by two functions: $b$ and $s$. These authors show how to compute these functions  so that the scaled expected length is minimized under model ${\cal M}_1$, subject to the constraints that (a) the coverage probability of this confidence interval never falls below $1-\alpha$, (b) the maximum scaled expected length under model ${\cal M}_2$ is not too large and (c) as the data becomes less consistent with the model  ${\cal M}_1$, the scaled expected length approaches 1. They found that (to within computational accuracy) the coverage probability of the resulting confidence interval is $1-\alpha$ throughout the parameter space.   These Kabaila and Giri optimized confidence intervals then provide an upper bound on the performance of the MATA for any reasonable weight function.  Knowing how the Kabaila and Giri optimized confidence intervals perform enables us to formulate four key scenarios within which we structure our examination of the performance of the MATA with specified weight.   These scenarios are used to evaluate the MATA for a broad class of weights
based on exponentiating criteria related to Mallows' $C_P$. Our results show that, while far from ideal, this MATA performs surprisingly well, provided that we choose a member of this class that does not put too much weight on the simpler model ${\cal M}_1$.

\section{The MATA with a general weight function}

\subsection{The models}

Suppose that the model ${\cal M}_2$ is given by
\[
\YY = \XX \be + \vare,
\]
where $\YY$ is a random $n$-vector of responses, $\XX$ is a known
$n\times p$ model matrix with $p$ linearly independent columns, $\be$ is
an unknown $p$-vector parameter  and $\vare \sim \text{N}(\zer, \sigma^2 \II_n)$,
with $\sigma^2$ an unknown positive parameter. We write $m=n-p$ throughout the paper.  Suppose that we are interested in making inference about the parameter $\theta = \ba^{\top} \be$,  where $\ba$ is a specified nonzero $p$-vector.  Suppose also that we define the parameter $\tau = \cc^{\top} \be - t$,  where $\cc$ is a specified nonzero $p$-vector that is linearly independent of $\ba$ and $t$ is a specified number.  The model ${\cal M}_1$ is ${\cal M}_2$ with $\tau=0$.

Let $\hat{\be}$ be the least squares estimator of $\be$ and let $\hat{\sigma}^2 = (\YY-\XX\hat{\be})^{\top}(\YY-\XX\hat{\be})/m$
be the usual unbiased estimator of $\sigma^2$.  Set $\hat{\theta} = \ba^{\top} \hat{\be}$ and $\hat{\tau} = \cc^{\top} \hat{\be} - t$. Define $v_{\theta} = \ba^{\top}(\XX^{\top}\XX)^{-1}\ba$ and $v_{\tau} = \cc^{\top}(\XX^{\top}\XX)^{-1}\cc$.  Then two important quantities are the known correlation $\rho=\ba^{\top}(\XX^{\top}\XX)^{-1}\cc/(v_{\theta}v_{\tau})^{1/2}$ between $\hat{\theta}$ and $\hat{\tau}$ and the scaled unknown parameter $\gamma = \tau \big/\big(\sigma v_{\tau}^{1/2} \big)$.  We denote the estimator of $\gamma$ by $\hat{\gamma} = \hat{\tau} \big/\big(\hat{\sigma} v_{\tau}^{1/2} \big)$.


\subsection{The MATA}

The MATA is obtained by averaging the equations defining the tail area confidence
intervals under the models $\mathcal{M}_2$ and $\mathcal{M}_1$.  Suppose that
the weight function $w: [0, \infty) \rightarrow [0,1]$ is a decreasing continuous function, such that $w(z)$ approaches 0 as $z \rightarrow \infty$.
Any reasonable weight function must be of this form.
For each $a$ and $\gamma \in \mathbb{R}$, define
\begin{equation*}
k(a,\gamma)
= w(\gamma^2)
\, G_{m+1} \left\{\left( \frac{m + 1}{m + \gamma^2}\right)^{1/2}  \frac{a - \rho \, \gamma}{(1 - \rho^2)^{1/2}}  \right\}
 + \big\{1 - w(\gamma^2) \big\} \, G_m (a),
\end{equation*}
where $G_m$ denotes the distribution function of the Student t distribution with $m$ degrees of freedom.
The MATA,
with nominal coverage $1-\alpha$, is $ \left[\hat{\theta}_{\ell}, \,  \hat{\theta}_u \right]$, where
$\hat{\theta}_{l}$ and $\hat{\theta}_{u}$ are the solutions for $\theta$ of
\[
k \left\{(\hat{\theta}-\theta)/(\hat{\sigma}v_{\theta}^{1/2}), \hat{\gamma}\right\} = 1-\alpha/2 \ \  \mbox{ and } \ \ k \left\{(\hat{\theta}-\theta)/( \hat{\sigma}v_{\tau}^{1/2}), \hat{\gamma} \right\} = \alpha/2,
\]
respectively. 
Equivalently, if we let $a_{\ell}(\gamma)$ and $a_u(\gamma)$ be the solutions for $a$ of $k(a,\gamma) = 1 - \alpha/2$ and $k(a,\gamma) = \alpha/2$, respectively, then
the endpoints of the MATA 
are given by
\begin{align}
\label{alt_expr_MATA}
\begin{split}
\hat{\theta}_{\ell} &= \hat{\theta} - v_{\theta}^{1/2} \hat{\sigma} \, a_{\ell}(\hat{\gamma})
\\
\hat{\theta}_u &= \hat{\theta} - v_{\theta}^{1/2} \hat{\sigma} \, a_u(\hat{\gamma}).
\end{split}
\end{align}
The notation is slightly different from that used in Kabaila, Welsh and Abeysekara (2015), but the interval is the same.

\subsection{The Kabaila and Giri optimized confidence \newline
intervals}

The MATA, with nominal coverage $1-\alpha$, is a member of the class of
confidence intervals $J(b,s)$ defined by
Kabaila and Giri  (2009).  To see this, note that the centre and half-width of the MATA (\ref{alt_expr_MATA}) are $\hat{\theta} - v_{\theta}^{1/2} \hat{\sigma} \, \big\{a_{\ell}(\hat{\gamma}) + a_u(\hat{\gamma}) \big\}/2$
and $v_{\theta}^{1/2} \hat{\sigma} \, \big\{a_{\ell}(\hat{\gamma}) - a_u(\hat{\gamma}) \big\}/2$, respectively.
As shown in Appendix A, $a_{\ell}(\gamma) + a_u(\gamma)$ and $a_{\ell}(\gamma) - a_u(\gamma)$ are odd and even functions of $\gamma$, respectively, and $a_{\ell}(\gamma) + a_u(\gamma)$ and $a_{\ell}(\gamma) - a_u(\gamma)$ approach 0 and $2 G_m^{-1}(1 - \alpha/2)$, respectively, as $\gamma \rightarrow \infty$. Now define the functions $b$ and $s$ by $b(\gamma) = \{a_{\ell}(\gamma) + a_u(\gamma)\}/2$ and
$s(\gamma) = \{a_{\ell}(\gamma) - a_u(\gamma)\}/2$. Then the MATA,
with nominal coverage $1-\alpha$, can be written as
\begin{equation*}
\Big [ \hat{\theta} - v_{\theta}^{1/2} \hat{\sigma} \, b(\hat{\gamma}) - v_{\theta}^{1/2} \hat{\sigma} \, s(|\hat{\gamma}|), \, \,  \hat{\theta} - v_{\theta}^{1/2} \hat{\sigma} \, b(\hat{\gamma}) + v_{\theta}^{1/2} \hat{\sigma} \, s(|\hat{\gamma}|)  \Big],
\end{equation*}
which is of the form $J(b,s)$ considered by Kabaila and Giri (2009).  As proved by these authors, for given $1-\alpha$ and given functions $b$ and $s$, the coverage probability and the scaled expected length of $J(b,s)$ are functions  of the known quantities $(m, \rho)$ and the unknown parameter $\gamma$.
Since Kabaila and Giri (2009) optimized the choice of $b$ and $s$ separately for each given value of $(m, \rho)$,
we cannot expect that, for any given weight function $w$, the MATA
will perform better than the optimized confidence interval of Kabaila and Giri (2009).


\section{Weight functions based on Mallows' $\boldsymbol{C_P}$}

For the models ${\cal M}_2$ and ${\cal M}_1$ the Generalized  Information Criteria (GIC; Nishii, 1984, Rao and Wu, 1989) are
\[
\mbox{GIC}_2 = n\log\{m \hat{\sigma}^2/n\} + dp
\]
and
\[
\mbox{GIC}_1 = n\log[\{(\hat{\tau}^2/v_{\tau}) + m \hat{\sigma}^2 \}/n] + d(p-1),
\]
respectively, where $d$ is a specified nonnegative number.   The choices $d=2$ and $d = \log(n)$ yield AIC and BIC, respectively.  The corresponding weight function is
\begin{eqnarray*}
w_*\big(\hat{\gamma}^2; d \big) &=& \frac{\exp\big\{-\frac{1}{2}(\mbox{GIC}_1 - \mbox{GIC}_{\min})\big\}}{\exp\big\{-\frac{1}{2}(\mbox{GIC}_1 - \mbox{GIC}_{\min})\big\} + \exp\big\{-\frac{1}{2}(\mbox{GIC}_2 - \mbox{GIC}_{\min})\big\}} \\
&=& \frac{1}{1 + \Big\{1+\hat{\gamma}^2/m \Big\}^{n/2} \exp(-d/2)},
\end{eqnarray*}
where $\mbox{GIC}_{\min} = \min(\mbox{GIC}_1, \mbox{GIC}_2)$.

We now motivate the use of this weight function, with the power $n/2$ replaced by $m/2$. Define the criteria
\begin{equation*}
\mbox{MIC}_2 = m\log\{m\hat{\sigma}^2/n\} + dp
\end{equation*}
and
\begin{equation*}
\mbox{MIC}_1 = m\log[\{(\hat{\tau}^2/v_{\tau}) + m\hat{\sigma}^2 \}/n] + d(p-1),
\end{equation*}
for the models ${\cal M}_2$ and ${\cal M}_1$, respectively.   As we show in Appendix C, choosing ${\cal M}_2$ if and only if (Mallows' $C_P$ for ${\cal M}_2$) $\le$ (Mallows $C_P$ for ${\cal M}_1$) is equivalent to  choosing ${\cal M}_2$ if and only if $\mbox{MIC}_2 \le \mbox{MIC}_1$, for $d = m \log(1 + (2/m))$.   The criteria $\mbox{MIC}_2$  and $\mbox{MIC}_1$ lead to the weight function
\begin{align*}
\label{eq:weight}
w(\hat{\gamma}^2; d)
&= \frac{\exp\big\{-\frac{1}{2}(\mbox{MIC}_1 - \mbox{MIC}_{\min})\big\}}{\exp\big\{-\frac{1}{2}(\mbox{MIC}_1 - \mbox{MIC}_{\min})\big\} + \exp\big\{-\frac{1}{2}(\mbox{MIC}_2 - \mbox{MIC}_{\min})\big\}} \\
&= \frac{1}{1 + \Big(1+\hat{\gamma}^2/m\Big)^{m/2} \exp(-d/2)},
\end{align*}
where $\mbox{MIC}_{\min} = \min(\mbox{MIC}_1, \mbox{MIC}_2)$.
The criteria $\mbox{MIC}_2$  and $\mbox{MIC}_1$ are close to the criteria $\mbox{GIC}_2$  and $\mbox{GIC}_1$, respectively, for $p$ fixed and $n$ large, but replacing $n/2$ by $m/2$ achieves a useful simplification; the coverage probability and scaled expected length of the MATA, with nominal coverage $1-\alpha$, are determined by the known quantities $(d, m, \rho)$ and the
unknown parameter $\gamma$ (as in Kabaila and Giri, 2009) rather than by $(d, n, p, \rho)$ and $\gamma$ (as shown in Theorems 1 and 2 of Kabaila, Welsh and Abeysekera, 2015). The reduction from the 4 known quantities $(d, n, p, \rho)$ to the 3 known quantities $(d, m, \rho)$ represents a considerable gain in simplicity.

As also noted in Appendix C, using the MIC to choose between the models ${\cal M}_1$ and ${\cal M}_2$
is equivalent to testing the null hypothesis $\tau =0$ (i.e. ${\cal M}_1$) against the alternative hypothesis $\tau \ne 0$
with level of significance
\begin{equation*}
2\left(1-G_{m}\left[m^{1/2}\left\{\exp\left(\frac{d}{m}\right) - 1\right\}^{1/2}\right]\right).
\end{equation*}
Large values of $d$ correspond to small values of this level of significance and so can be interpreted as putting more weight on the simpler model ${\cal  M}_1$.

The interpretation of $d$ is quite different in the model selection and model averaging contexts. In the model selection context, setting
$d = 0$ means that there is no penalty on the number of parameters and so, for the models ${\cal M}_1$ and ${\cal M}_2$,
we always choose model ${\cal M}_2$. By contrast, in the model averaging context, setting $d = 0$ leads to
$w(\widehat{\gamma}^2; 0)=1 \big/ \big\{1 + (1+\widehat{\gamma}^2/m)^{m/2} \big\}$
 so we still average over the two models.


\section{How well can we expect MATA to perform?}

The performance of the Kabaila and Giri (2009, 2013) optimized confidence interval relative to the standard $1-\alpha$ Student t confidence interval under model ${\cal M}_2$ can be described under four different scenarios defined by the values of $m$ and $\rho$, as set out in Table \ref{performance_KG_interval}.  The MATA cannot perform better than the Kabaila and Giri optimized confidence interval so, for each of the four scenarios, we compare the coverage and scaled expected length properties of the MATA against the best we can hope for from the Kabaila and Giri optimized confidence interval.  Details of the methods used to compute the coverage and scaled expected length of the MATA are given in Appendix A.

\smallskip

\begin{table}[h]
\begin{center}
\begin{tabular}{|c|c|c|}
\hline
  & $|\rho|$ is small & $|\rho|$ is not small
\\
  \hline
  $m$ is not small & \textbf{Scenario 1} & \textbf{Scenario 2}
  \\
                &  Cannot do better than the          &  Some improvement over the
  \\
               &   standard $1-\alpha$ t interval           & standard $1-\alpha$ t interval
  \\
               &   for $\theta$           & for $\theta$
  \\
  \hline
  $m$ is small& \textbf{Scenario 3} & \textbf{Scenario 4}
  \\
  (i.e. 1, 2 or 3) & Some improvement over the          &  Maximum improvement over
  \\
              &   standard $1-\alpha$ t interval           & standard $1-\alpha$ t interval
  \\
               &   for $\theta$           & for $\theta$
  \\
  \hline
\end{tabular}
\end{center}
\caption{Performance of the optimized confidence interval of Kabaila and Giri (2009) for various values of the known quantities
$m$ and $\rho$.}
\label{performance_KG_interval}
\end{table}

In all our numerical work, we fix the nominal coverage $1-\alpha=0.95$ and vary the values of $ \rho$ and $m$ according to the different scenarios (other than the first which we are able to treat theoretically).  As in Kabaila, Welsh and Abeysekera (2015), for each scenario we can compute and plot the coverage and scaled expected length of the MATA for fixed $d$ as functions of $\gamma$.   Typically, for fixed $d$, the coverage is greater than $0.95$, decreases to a minimum value and then increases to asymptote at $0.95$ as $\gamma$ increases; the scaled expected length is less than one for $\gamma=0$, increases to a maximum and then decreases to an asymptote (often greater than one) as $\gamma$ increases.  Some examples of these kinds of figures are included in Kabaila, Welsh and Abeysekera (2015).  For our present purposes, it is useful to summarise the above results over different choices of $d\in [0,8]$ by presenting  the minimum coverage and maximum scaled expected length over $\gamma$ for each fixed $d$
and the scaled expected length at $\gamma=0$  for each fixed $d$
as functions of $d$.   We make these quantities positive with a baseline value of zero by computing
the coverage loss (\textsl{cov loss}), scaled expected length loss (\textsl{sel loss}) and the scaled expected length gain
(\textsl{sel gain}), defined as follows
\begin{align*}
\textsl{cov loss} &= (1- \alpha) - \text{(minimum coverage)} \\
\textsl{sel loss} &= \text{(maximum scaled expected length)} - 1 \\
\textsl{sel gain} &= 1 - \text{(scaled expected length at $\gamma = 0$)}.
\end{align*}
Ideally, one would have a high \textsl{sel gain} together with small \textsl{cov loss} and \textsl{sel loss}.

\subsection{Scenario 1}

This scenario includes the cloud seeding example in Section 3 of Kabaila, Welsh and Abeysekera (2015) where $\rho = 0.2472$ and $m = 11$. For \textbf{Scenario 1} we cannot expect the MATA, with nominal coverage $1-\alpha$, to perform any better than the standard $1-\alpha$ confidence interval for $\theta$ based on model ${\cal M}_2$ so the best hope is that MATA recovers the standard $1-\alpha$ confidence interval for $\theta$ based on model ${\cal M}_2$.  That is, that
\begin{equation}
\label{est_eqn_BS1}
k(a, \hat{\gamma}) \approx G_m (a).
\end{equation}
Indeed, if $|\rho|$ is small, then
\begin{align*}
k(a, \hat{\gamma})
&\approx w( \hat{\gamma}^2)
G_{m+1} \left\{ \left(\frac{m + 1}{m + \hat{\gamma}^2} \right)^{1/2} a\right\}
 + \{1 - w( \hat{\gamma}^2) \} G_m (a).
\end{align*}
If, in addition, $m$ is not small, then $G_{m+1} \approx G_m$ and, since $w$ is a decreasing continuous function, such that $w(z)$ approaches 0 as $z \rightarrow \infty$,
\begin{align*}
k (a, \hat{\gamma})
&\approx w( \hat{\gamma}^2)
G_m (a)
 + \{1 - w( \hat{\gamma}^2) \} G_m (a) =  G_m (a),
\end{align*}
so (\ref{est_eqn_BS1}) holds. In Scenario 1, for given $\rho$ and $m$, as $d$ increases MATA is less likely to recover the standard $1-\alpha$ confidence interval for $\theta$ based on model ${\cal M}_2$. So, in Scenario 1 a good choice of $d$ is 0.

\subsection{Scenario 2}

\begin{figure}[h]
\begin{center}
\includegraphics[scale=0.8]{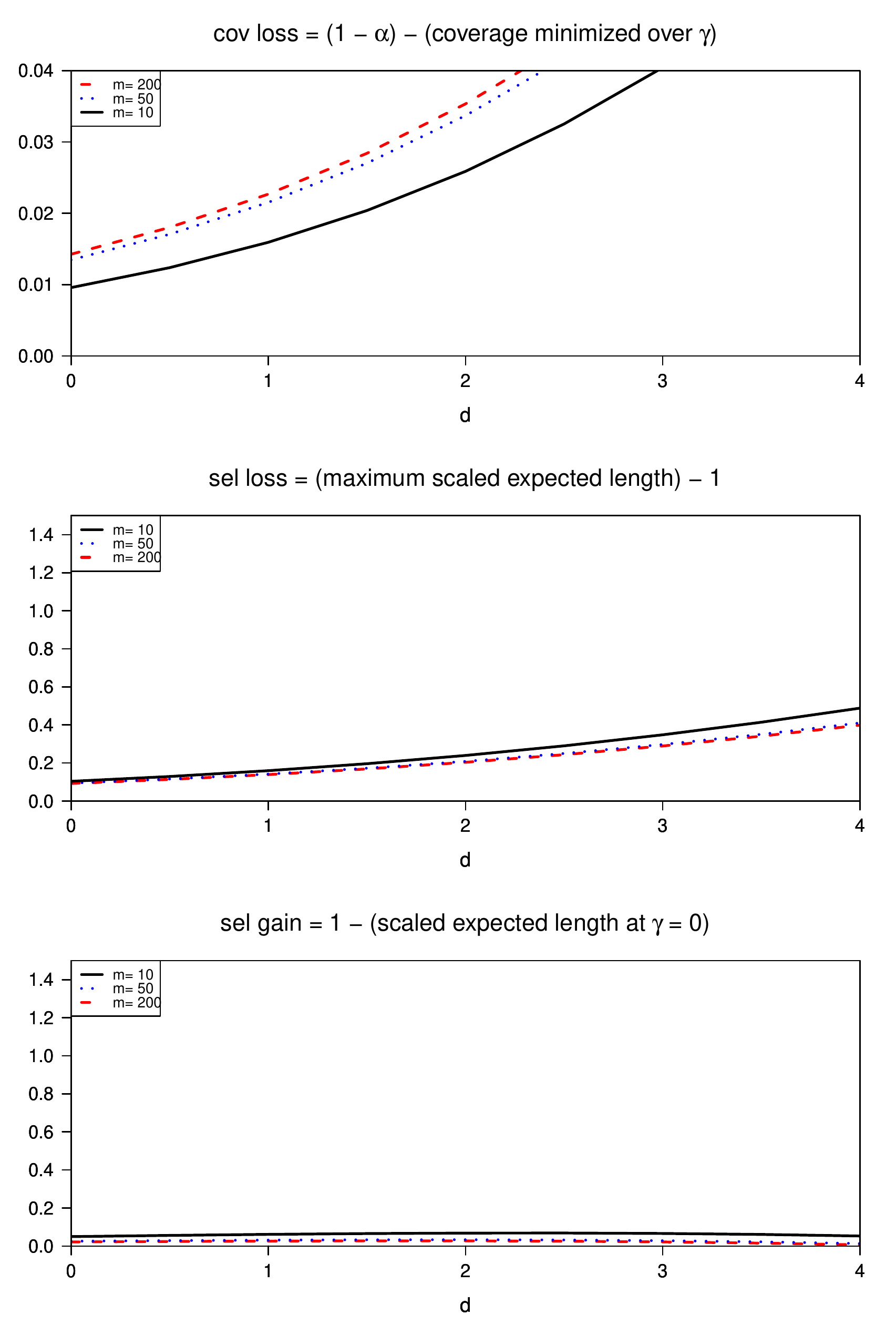}
\caption{\small{Graphs of the {\sl cov loss}, {\sl sel loss} and {\sl sel gain} of the MATA, with weight determined by $d$ and nominal coverage 95\%.
Here, $\rho = \text{Corr}(\hat{\theta}, \hat{\tau}) = 0.8$ and $m \in \{10, 50, 200 \}$.
}}
\label{fig1}
\end{center}
\end{figure}

Graphs of \textsl{cov loss}, \textsl{sel loss} and \textsl{sel gain} as functions of $d\in [0,4]$ for $\rho=0.8$ and $m \in \{10, 50, 200\}$ are shown in Figure \ref{fig1}.  These functions are displayed only for $d\in [0,4]$ because \textsl{cov loss} and \textsl{sel loss} are
both increasing functions and \textsl{sel gain} is a decreasing function of $d$ in $[4,8]$. In other words, when searching for a good value of $d$ there is no point in considering values of $d$ in $[4,8]$. For $m = 10$, as we increase $d$ from 0 to 4, the \textsl{cov loss} is an increasing function of $d$, \textsl{sel gain} changes slowly and \textsl{sel loss} increases. In this circumstance, a good choice of $d$ is 0. Similarly, for $m = 50$ and $m = 200$, a good choice of $d$ is 0.
These results show that there is not much gain from using the MATA in Scenario 2.

\subsection{Scenario 3}

\begin{figure}[h]
\begin{center}
\includegraphics[scale=0.8]{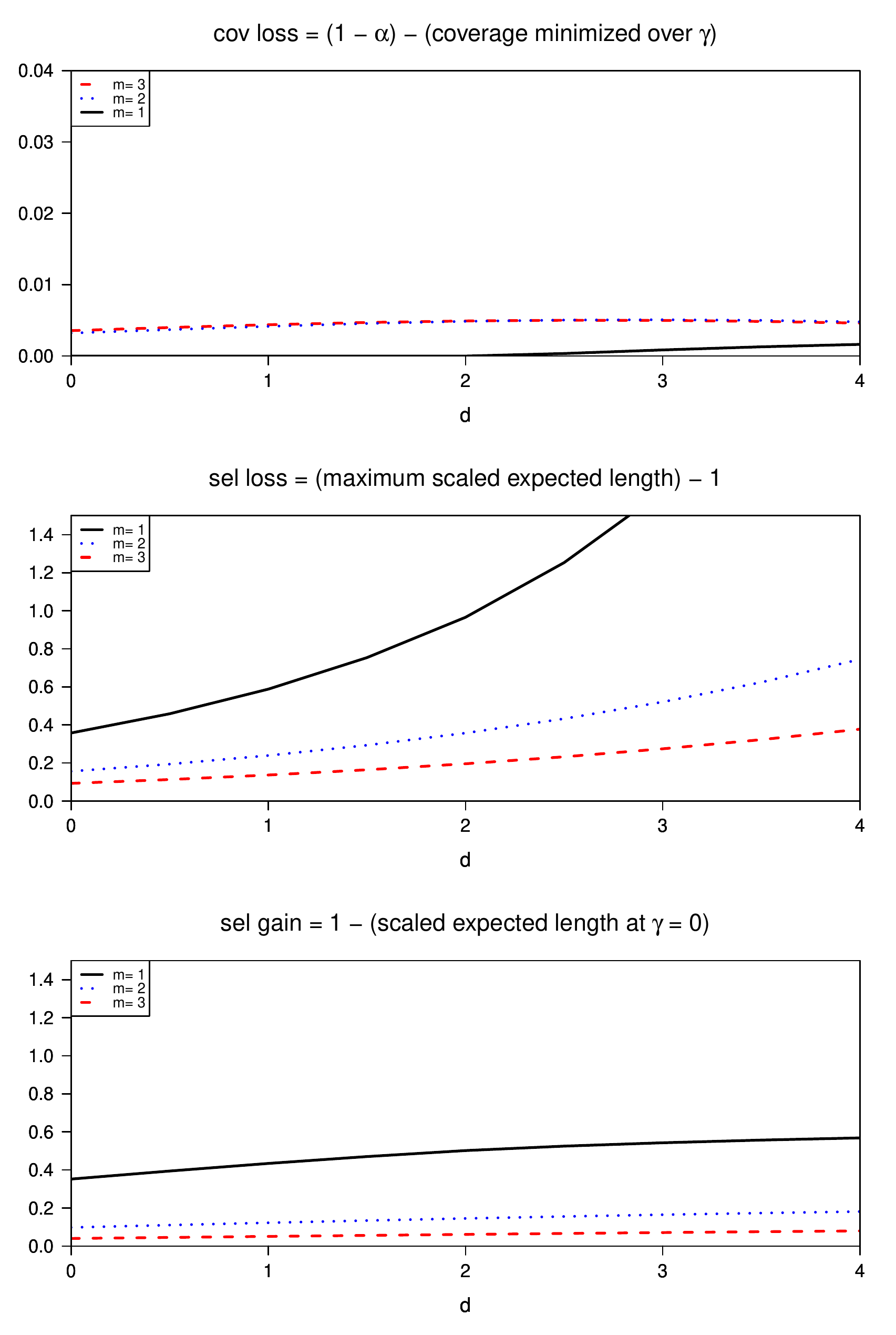}
\caption{\small{Graphs of the {\sl cov loss}, {\sl sel loss} and {\sl sel gain} of the MATA, with weight determined by $d$ and nominal coverage 95\%.
Here, $\rho = \text{Corr}(\hat{\theta}, \hat{\tau}) = 0$ and $m \in \{1, 2, 3\}$.
}}
\label{fig2}
\end{center}
\end{figure}

Graphs of \textsl{cov loss}, \textsl{sel loss} and \textsl{sel gain} as functions of $d\in [0,4]$ for $\rho=0$ and $m \in \{1,2,3\}$ are shown in Figure \ref{fig2}. For $m=2$ and $m=3$, the \textsl{cov loss} is a decreasing function of $d$ in $[4,8]$. Also, for $m = 1$, the \textsl{cov loss} is
initially an increasing function and then a decreasing function of $d$ in $[4,8]$. However, the \textsl{cov loss} remains small for all $d$ in $[0,8]$
and \textsl{sel loss} is an increasing function of $d$ in $[0,8]$. Therefore, when searching for a good value of $d$, it seems reasonable to
restrict consideration to values of $d$ in $[4,8]$.
For $m = 1$, as we increase $d$ from 0 to 4, the \textsl{sel gain} increases slowly and \textsl{sel loss} increases much more quickly. In this circumstance, a good choice of $d$ is 0. Similarly, for $m = 2$ and $m = 3$, a good choice of $d$ is 0.
In Scenario 3, there is a small gain from using the MATA, provided that we choose $d$ near 0.

\subsection{Scenario 4}

\begin{figure}[h]
\begin{center}
\includegraphics[scale=0.8]{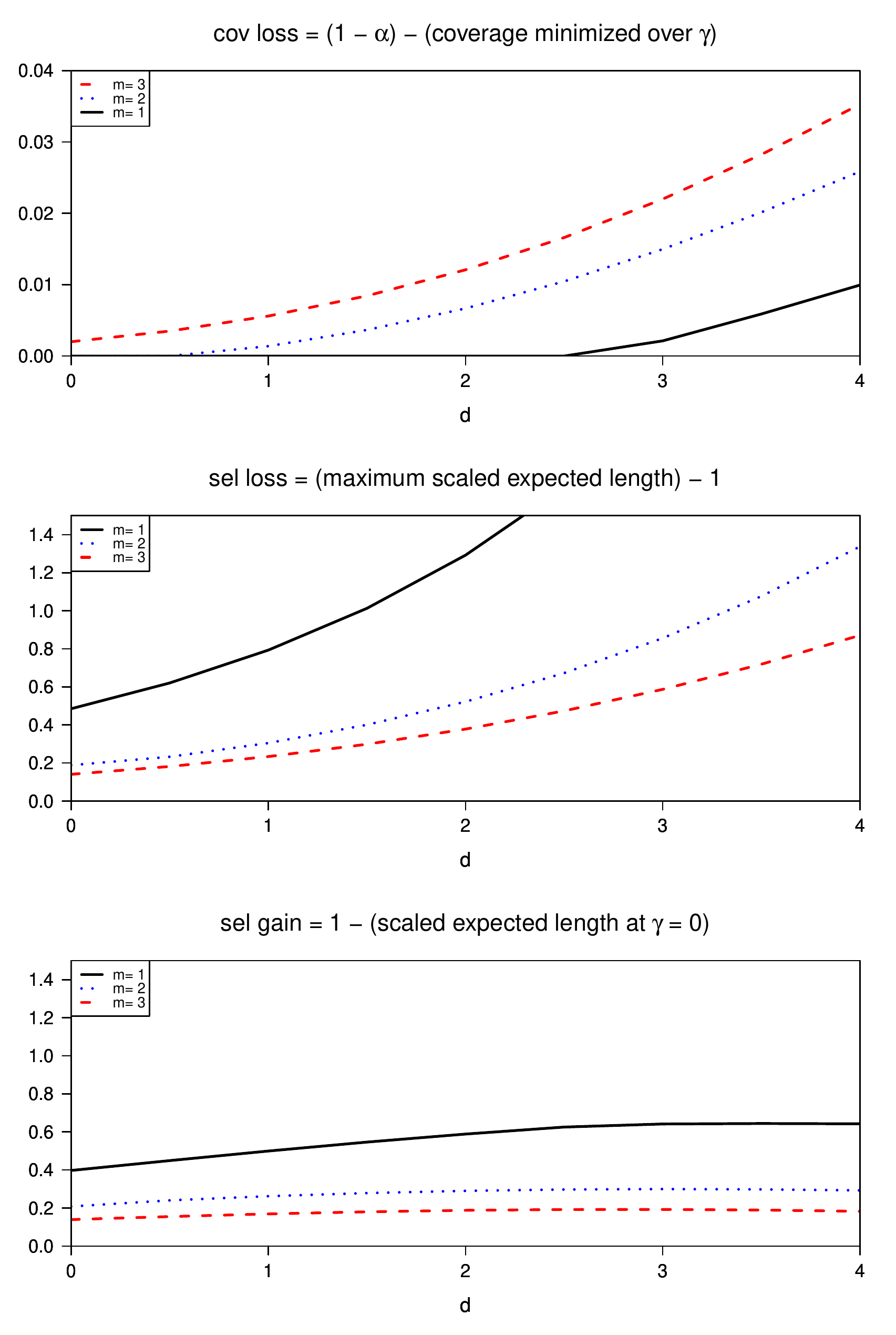}
\caption{\small{Graphs of the {\sl cov loss}, {\sl sel loss} and {\sl sel gain} of the MATA, with weight determined by $d$ and nominal coverage 95\%.
Here, $\rho = \text{Corr}(\hat{\theta}, \hat{\tau}) = 0.8$ and $m \in \{1, 2, 3 \}$.
}}
\label{fig3}
\end{center}
\end{figure}

Graphs of \textsl{cov loss}, \textsl{sel loss} and \textsl{sel gain} as functions of $d\in [0,4]$ for $\rho=0.8$ and $m \in \{1,2,3\}$ are shown in Figure \ref{fig3}.  These functions are displayed only for $d\in [0,4]$ because \textsl{cov loss} and \textsl{sel loss} are
both increasing functions and \textsl{sel gain} is a decreasing function of $d$ in $[4,8]$. In other words, when searching for a good value of $d$ there is no point in considering values of $d$ in $[4,8]$. For $m = 1$, as we increase $d$ from 0 to 4, the \textsl{cov loss} is a nondecreasing function of $d$, \textsl{sel gain} increases slowly and \textsl{sel loss} increases much more quickly. In this circumstance, a good choice of $d$ is 0. Similarly, for $m = 2$ and $m = 3$, a good choice of $d$ is 0.

\section{Conclusion}

We have examined the exact coverage and scaled expected length of the MATA for a parameter $\theta$, with nominal level $1-\alpha$, in a particular simple situation in which there are two linear regression models (differing in only a single parameter $\tau$) to average over.
For weight functions based on Mallows' $C_P$ (specified by $d \in [0, 8]$),
we showed that both the coverage and the scaled expected length depend on $m=n-p$, the correlation $\rho$ between the least squares estimators $\hat{\theta}$ and $\hat{\tau}$, and the unknown parameter $\gamma=\tau \big/ \big(\sigma v_{\tau}^{1/2} \big)$.
We assess the MATA according to two losses, using the minimum coverage and the maximum scaled expected length over $\gamma$
and a gain, using the scaled expected length at $\gamma = 0$ i.e. when the simpler model ${\cal M}_1$ is true.
The results show that, although the MATA is far from ideal, it performs surprisingly well, provided that we do
not put too much weight on the model ${\cal M}_1$.


\section*{References}






\rf Buckland, S.T., Burnham, K.P. and Augustin, N.H. (1997). Model selection: an integral
part of inference. \textit{Biometrics}, \textbf{53}, 603--618.

\smallskip

\rf Kabaila, P. and Giri, K. (2009). Confidence intervals in regression utilizing uncertain prior information. \textit{Journal of Statistical Planning and Inference}, \textbf{139}, 3419--3429.

\smallskip

\rf Kabaila, P. and Giri, K. (2013). Further properties of frequentist confidence intervals in regression that utilize uncertain prior information. \textit{Australian \& New Zealand Journal of Statistics}, \textbf{55}, 259--270.

\smallskip

\rf Kabaila, P., Welsh, A.H. and Abeysekera, W. (2015). Model averaged confidence intervals. To appear in \textit{Scandinavian Journal of Statistics}.

%
%





\smallskip
\rf Nishii, R. (1984). Asymptotic properties of criteria for selection of variables in multiple regression. \textit{The Annals of Statistics}, \textbf{12}, 758--765.

%
%



\smallskip
\rf Rao, C.R.  and Wu, Y. (1989). A strongly consistent procedure for model selection in regression problems. \textit{Biometrika}, \textbf{76}, 369--374.

\smallskip

\rf Turek, D. and Fletcher, D. (2012). Model-averaged Wald confidence intervals. \textit{Computational Statistics and Data Analysis}, \textbf{56},
2809--2815.





\bigskip


\section*{Appendix A: The MATA is in the class of \newline confidence intervals $\boldsymbol{J(b,s)}$}


We show that $a_{\ell}(\gamma) + a_u(\gamma)$ and $a_{\ell}(\gamma) - a_u(\gamma)$ are odd and even functions of $\gamma$, respectively, as follows. Recall that
$a_u(-\gamma)$ is the solution for $a$ of $k(a,-\gamma) = \alpha/2$ i.e.
$a_u(-\gamma)$ is the solution for $a$ of
\begin{equation*}
 w(\gamma^2)
\, G_{m+1} \left\{\left( \frac{m + 1}{m + \gamma^2}\right)^{1/2}  \frac{a + \rho \, \gamma}{(1 - \rho^2)^{1/2}}  \right\}
 + \big\{1 - w(\gamma^2) \big\} \, G_m (a) =  \frac{\alpha}{2}.
\end{equation*}
Using the fact that the probability density function of a Student t distribution is an even function, we can show that
\begin{align*}
w(\gamma^2)&
\, G_{m+1} \left\{\left( \frac{m + 1}{m + \gamma^2}\right)^{1/2}  \frac{a_u(-\gamma) + \rho \, \gamma}{(1 - \rho^2)^{1/2}}  \right\}
 + \big\{1 - w(\gamma^2) \big\} \, G_m \{a_u(-\gamma)\}\\
 &= w(\gamma^2)\,  \left[ 1-G_{m+1} \left\{-\left( \frac{m + 1}{m + \gamma^2}\right)^{1/2}  \frac{a_u(-\gamma) + \rho \, \gamma}{(1 - \rho^2)^{1/2}}  \right\}\right]\\
 &\qquad+ \big\{1 - w(\gamma^2) \big\} \, [1-G_m \{- a_u(-\gamma)\}]\\
 %
%
&= 1\,- w(\gamma^2) G_{m+1} \left\{\left( \frac{m + 1}{m + \gamma^2}\right)^{1/2}  \frac{-a_u(-\gamma) - \rho \, \gamma}{(1 - \rho^2)^{1/2}}  \right\}\\
& \qquad - \{1- w(\gamma^2)\} \, G_m \{- a_u(-\gamma)\}
\end{align*}
so $a_u(-\gamma) = - a_{\ell}(\gamma)$. Thus  $a_{\ell}(-\gamma) = - a_u(\gamma)$ and it follows that
$a_{\ell}(\gamma) + a_u(\gamma)$ and $a_{\ell}(\gamma) - a_u(\gamma)$ are odd and even functions of $\gamma$, respectively.

We now show that $a_{\ell}(\gamma) + a_u(\gamma)$ and $a_{\ell}(\gamma) - a_u(\gamma)$ approach 0 and $2 G_m^{-1}(1 - \alpha/2)$, respectively, as $\gamma \rightarrow \infty$.
As $\gamma \rightarrow \infty$,
$k(a,\gamma)$ approaches $G_m (a)$. Therefore $a_{\ell}(\gamma)$ and $a_u(\gamma)$ approach the solutions for $a$ of $G_m (a) = 1 - \alpha/2$
and $G_m (a) = \alpha/2$, respectively. Thus $a_{\ell}(\gamma) + a_u(\gamma)$ and $a_{\ell}(\gamma) - a_u(\gamma)$ approach 0 and
$2 G_m^{-1}(1 - \alpha/2)$, respectively,
as $\gamma \rightarrow \infty$.

\newpage

\section*{Appendix B: Computational methods}

The computation of the coverage and the scaled expected length of the MATA are implemented using Theorems 1 and 2 of Kabaila, Welsh and Abeysekera (2015).  In the following two sections, we establish results which are needed to support the computations.

\subsection*{Truncation of integrals}

To compute the coverage and the scaled expected length of the MATA, we need to truncate the integrals in Theorems 1 and 2 of
Kabaila, Welsh and Abeysekera (2015).  The coverage probability integral in Theorem 1 is relatively easy to handle because cumulative distribution functions are bounded.  If we write the integrand in Theorem 1 as $\Delta_m(x,y, \rho, \gamma)$, we have
\begin{eqnarray*}
\lefteqn{\left|\Pr(\hat{\theta}_{\ell} \le \theta \le \hat{\theta}_u) -
\int_{y_{\ell}}^{y_u} \int_{x_{\ell}}^{x_u} \Delta_m(x,y, \rho, \gamma) dx dy \right|}\\
&&  \le \left|\int_{0}^{\infty} \int_{-\infty}^{\infty} \Delta_m(x,y, \rho, \gamma) dx dy
- \int_{y_{\ell}}^{y_u} \int_{-\infty}^{\infty} \Delta_m(x,y, \rho, \gamma) dx dy \right|\\
&& + \left|\int_{y_{\ell}}^{y_u} \int_{-\infty}^{\infty} \Delta_m(x,y, \rho, \gamma) dx dy -
\int_{y_{\ell}}^{y_u} \int_{x_{\ell}}^{x_u} \Delta_m(x,y, \rho, \gamma) dx dy \right|\\
&&  \le \int_{y_u}^{\infty} \int_{-\infty}^{\infty} |\Delta_m(x,y, \rho, \gamma)| dx dy
+  \int_0^{y_{\ell}} \int_{-\infty}^{\infty} |\Delta_m(x,y, \rho, \gamma)| dx dy\\
&& +\int_{y_{\ell}}^{y_u} \int_{x_u}^{\infty} |\Delta_m(x,y, \rho, \gamma)| dx dy +
\int_{y_{\ell}}^{y_u} \int_{-\infty}^{x_{\ell}} |\Delta_m(x,y, \rho, \gamma)| dx dy.
\end{eqnarray*}
Now
$|\Delta_m(x,y, \rho, \gamma) | \le \phi(x-\gamma) g_m(y)$ so
\[
\int_{y_u}^{\infty} \int_{-\infty}^{\infty} |\Delta_m(x,y, \rho, \gamma)| dx dy \le 1-G_m(y_u),
\]
\[
\int_0^{y_{\ell}} \int_{-\infty}^{\infty} |\Delta_m(x,y, \rho, \gamma)| dx dy \le G_m(y_{\ell}),
\]
\[
\int_{y_{\ell}}^{y_u} \int_{x_u}^{\infty} |\Delta_m(x,y, \rho, \gamma)| dx dy \le \{G_m(y_u) - G_m(y_{\ell})\}\{1-\Phi(x_u - \gamma)\}
\]
and
\[
\int_{y_{\ell}}^{y_u} \int_{-\infty}^{x_{\ell}} |\Delta_m(x,y, \rho, \gamma)| dx dy \le \{G_m(y_u) - G_m(y_{\ell})\}\Phi(x_{\ell} - \gamma).
\]
For any given small positive number $\epsilon$, set $y_{\ell} = G_m^{-1}(\epsilon/4)$ and $y_u = G_m^{-1}(1-\epsilon/4)$ so the first two terms are both less than or equal to $\epsilon/4$. If we also set $x_{\ell} = \Phi^{-1}(\epsilon/4) + \gamma$ and $x_u= \Phi^{-1}(1-\epsilon/4) + \gamma$, the last two terms are both less than or equal to $(1-\epsilon/2)\epsilon/4 \le \epsilon/4$  and the sum of all the terms is less than or equal to $\epsilon$.  Thus the effect of the truncation can be made as small as we want.

The scaled expected length is more difficult to handle because the integrand is unbounded.  Nonetheless, $\delta_{1-\alpha/2}(x,y) - \delta_{\alpha/2}(x,y)$ is approximately constant in $x$ and linear in $y$ and we can use this approximation to similarly find truncation values for the integrals in Theorem 2  so that the effect of the truncation is arbitrarily small.

\subsection*{Computation of $\boldsymbol{\delta_u(x,y)}$}

The formulae for the coverage probability and the scaled expected length of the MATA, with nominal coverage probability
$1-\alpha$, given by Kabaila, Welsh and Abeysekera (2015) require the computation of $\delta_u(x,y)$, which is defined to be the solution
for $\delta$ of
\begin{equation*}
h(\delta, x, y) = u,
\end{equation*}
where $0 < u < 1$ and
\begin{equation}
\label{def_h}
h(\delta,x,y) = w(x^2/y^2) \, G_{m+1} ( r_1(\delta, x, y))
+ \{1-w(x^2/y^2)\} \, G_{m}(r_2(\delta, y)).
\end{equation}
The definitions of $r_1(\delta, x, y)$ and $r_2(\delta, y)$ are given on p.6 of Kabaila, Welsh and Abeysekera (2015).
The numerical computation of $\delta_u(x,y)$ typically requires the provision of an interval known to contain $\delta_u(x,y)$.
The following result provides just such an interval.

\medskip

\noindent \textbf{Result A1}. \ Let $\delta_u^{(1)}(x,y)$ denote the solution for $\delta$ of
\begin{equation*}
G_{m+1} ( r_1(\delta, x, y)) = u.
\end{equation*}
Also, let $\delta_u^{(2)}(y)$ denote the solution for $\delta$ of
\begin{equation*}
 G_{m}(r_2(\delta, y)) = u.
\end{equation*}
The following are explicit expressions for $\delta_u^{(1)}(x,y)$ and $\delta_u^{(2)}(y)$:
\begin{align*}
\delta_u^{(1)}(x,y)
&= \rho \, x + G_{m+1}^{-1}(u) \, y  \, \left( \frac{m + (x^2/y^2)}{m+1}  \right)^{1/2} (1 - \rho^2)^{1/2}
\\
\delta_u^{(2)}(y)
&= G_m^{-1}(u) \, y.
\end{align*}
Then
\begin{equation}
\label{interval_for_deltau}
\delta_u(x,y)
\in \left[ \min \big(\delta_u^{(1)}(x,y), \delta_u^{(2)}(y) \big),  \max \big(\delta_u^{(1)}(x,y), \delta_u^{(2)}(y) \big) \right]
\end{equation}
\medskip

\noindent \textbf{Proof} \ Suppose that $(x,y)$ is given. Since $r_1(\delta, x, y)$ is a continuous increasing function
of $\delta$, $G_{m+1} ( r_1(\delta, x, y))$ is also a continuous increasing function
of $\delta$.
Also, since $r_2(\delta, y)$ is a continuous increasing function
of $\delta$, $G_m ( r_2(\delta, y))$ is also a continuous increasing function
of $\delta$.
It follows from \eqref{def_h}
that $h(\delta, x, y)$ is a continuous increasing function
of $\delta$ and that
\begin{equation*}
\min \big(G_{m+1} ( r_1(\delta, x, y)),G_{m}(r_2(\delta, y)) \big)
\le h(\delta,x,y)
\le \max \big(G_{m+1} ( r_1(\delta, x, y)),G_{m}(r_2(\delta, y)) \big).
\end{equation*}
It follows from this that \eqref{interval_for_deltau} is satisfied.

\hfill $\qed$

\bigskip



\section*{Appendix C: Weights based on Mallows' $\boldsymbol{C_P}$}

Let $\text{RSS} = (\YY-\XX\hat{\be})^T(\YY-\XX\hat{\be}) = m\hat{\sigma}^2$.  Also, let $\text{RSS}_* = (\YY-\XX\hat{\be}_*)^T(\YY-\XX\hat{\be}_*)$, where $\hat{\be}_*$ denotes the least squares estimator of $\be$ subject to the constraint $\tau=0$.  Note that $\text{RSS}_* = \hat{\tau}^2/v_{\tau} + m\hat{\sigma}^2$.
Mallows' $C_P$ for ${\cal M}_2$ is
\[
\frac{\text{RSS}}{\text{RSS}/m} - n + 2p = n-p-n+2p = p
\]
and Mallows' $C_P$ for ${\cal M}_1$ is
\[
\frac{\text{RSS}_*}{\text{RSS}/m} - n + 2(p-1) = \frac{(\hat{\tau}^2/v_{\tau})  + m\hat{\sigma}^2}{\hat{\sigma}^2} - n+2p-1.
\]

\noindent \textbf{Result C1}.  Choosing ${\cal M}_2$ if and only if (Mallows' $C_P$ for ${\cal M}_2$) $\le$ (Mallows $C_P$ for ${\cal M}_1$) is equivalent to choosing
${\cal M}_2$ if and only if $\mbox{MIC}_2 \le \mbox{MIC}_1$, for $d = m \log(1 + (2/m))$.

\noindent \textbf{Proof}: The result follows from the fact that (Mallows' $C_P$ for ${\cal M}_2$) $\le$ (Mallows $C_P$ for ${\cal M}_1$) is equivalent to the following four statements
\begin{align*}
m + 2 &\le \frac{(\hat{\tau}^2/v_{\tau} ) + m\hat{\sigma}^2}{\hat{\sigma}^2}, \\
m\log(\hat{\sigma}^2/n)& + m\log(m+2)  \le  m\log[\{(\hat{\tau}^2/v_{\tau})  + m\hat{\sigma}^2\}/n],\\
m\log(\hat{\sigma}^2/n)& + m\log(m) + m\log(1+2/m) \le  m\log[\{(\hat{\tau}^2/v_{\tau})  + m\hat{\sigma}^2\}/n],\\
m\log(m\hat{\sigma}^2/n) & \le  m\log[\{(\hat{\tau}^2/v_{\tau})  + m\hat{\sigma}^2\}/n] - d,
\end{align*}
where $d = m \log(1 + (2/m))$, and the last of these is equivalent to $\mbox{MIC}_2 \le \mbox{MIC}_1$.

\noindent \textbf{Result C2}. Choosing ${\cal M}_2$ if and only if $\mbox{MIC}_2 \le \mbox{MIC}_1$ is equivalent to  testing the null hypothesis $\tau=0$ against the alternative hypothesis $\tau \ne 0$ with level of significance
\[
2\left(1-G_m\left[m\left\{\exp\left(\frac{d}{m} \right) - 1\right\}^{1/2} \right] \right).
\]

\noindent \textbf{Proof}: The null hypothesis $H_0:\tau=0$ corresponds to choosing $\mathcal{M}_1$ so the level of significance of the test is the probability of choosing $\mathcal{M}_2$ under the null hypothesis $H_0$ and
\begin{align*}
\Pr(\mbox{MIC}_1 &\ge \mbox{MIC}_2)\\
& = \Pr\left(m\log[\{(\hat{\tau}^2/v_{\tau})  + m\hat{\sigma}^2\}/n] - d \ge m\log(m\hat{\sigma}^2/n) \right)\\
&= \Pr \left\{\frac{(\hat{\tau}^2/v_{\tau} ) + m\hat{\sigma}^2}{m\hat{\sigma}^2} \ge \exp\left(\frac{d}{m} \right)\right\}\\
&= \Pr\left[\hat{\gamma}^2 \ge m\left\{\exp\left(\frac{d}{m} \right) - 1\right\}\right]\\
%
&=1- \Pr\left[-m^{1/2}\left\{\exp\left(\frac{d}{m} \right) - 1\right\}^{1/2}  \le \hat{\gamma} \le m^{1/2}\left\{\exp\left(\frac{d}{m} \right) - 1\right\}^{1/2} \right]\\
& = 1-G_m\left[m^{1/2} \left\{\exp\left(\frac{d}{m} \right) - 1\right\}^{1/2} \right] + G_m\left[-m^{1/2}\left\{\exp\left(\frac{d}{m} \right) - 1\right\}^{1/2} \right] \ \text{under}\ H_0\\
& = 2\left(1-G_m\left[m^{1/2} \left\{\exp\left(\frac{d}{m} \right) - 1\right\}^{1/2} \right] \right)  \ \text{under}\ H_0.
\end{align*}

\newpage

\FloatBarrier

\newpage

\FloatBarrier

\newpage

\FloatBarrier

\end{document}